\documentclass{emulateapj}

\begin{document}
\newcommand{\Lya}{{$Ly\alpha$}} 
\newcommand{\acsg}{{$g_{475}$}}
\newcommand{\acsi}{{$i_{775}$}}
\newcommand{\acsz}{{$z_{850}$}}
\newcommand{\nicJ}{{$J_{110}$}}
\newcommand{\nicH}{{$H_{160}$}}
\newcommand{\hst}{{\it HST}}

\title{An Overdensity of \boldmath $i$-dropouts Among A Population of Excess Field Objects in the Virgo Cluster\altaffilmark{1}}

\author{Haojing Yan\altaffilmark{2}, Nimish P. Hathi\altaffilmark{3} \& Rogier A. Windhorst\altaffilmark{4}}

\altaffiltext{1}{Based on observations made with the NASA/ESA Hubble Space
 Telescope, obtained at the Space Telescope Science Institute, which is 
 operated by the Association of Universities for Research in Astronomy, Inc.,
 under NASA contract NAS 5-26555. These observations are associated with 
 programs \#9780, 9488, 9575, 9584, and 9984}

\altaffiltext{2} {The Observatories of the Carnegie Institution of Washington,
813 Santa Barbara Street, Pasadena, CA 91101; yhj@ociw.edu}
\altaffiltext{3} {Department of Physics, Arizona State University, Tempe, AZ 85287}
\altaffiltext{4} {School of Earth \& Space Exploration, Arizona State University, Tempe, AZ 85287}

\begin{abstract}

   Using a set of deep imaging data obtained by the Advanced Camera for Surveys
(ACS) on the {\it Hubble Space Telescope} (\hst) shortly after its deployment,
Yan, Windhorst \& Cohen (2003) found a large number of F775W-band dropouts
($i$-dropouts), which are consistent with being galaxies at $z\approx 6$. The
surface density of $i$-dropouts thus derived, however, is an order of magnitude
higher than those subsequent studies found in other deep ACS fields, including
the {\it Hubble} Ultra-Deep Field (HUDF). Here we revisit this problem, using
both the existing data and the new data obtained for this purpose. We confirm
that the large overdensity of $i$-dropouts does exist in this field, and that
their optical-to-IR colors are similar to those in the HUDF. However, we have 
discovered that the $i$-dropout overdensity is accompanied with an even larger
excess of faint field objects in this region and its vicinity. This large
excess of field objects is most likely caused by the fact that we have resolved
the faint diffuse light extending from an interacting galaxy pair in the Virgo 
Cluster, M60/NGC4647, which lies several arcminutes away from the region where
the excess is found. The integrated light from the excess is a few percent of
the luminosity of the galaxy pair. We argue that this excess population is most
likely tidal ``debris'' and/or halo stars related to the galaxy pair rather
than to the Virgo Cluster in general. The $i$-dropouts in this field are within
the magnitude range where this excess of field objects occurs. The spatial 
distribution of the $i$-dropouts seems to follow the same gradient as the 
excess field population. This excess population is also red in color, and the
red wing of its color distribution continuously extends to the regime where the
$i$-dropouts reside. While we still cannot completely rule out the possibility
that the overdensity of $i$-dropouts might be a genuine large-scale structure
of galaxies at $z\approx 6$, we prefer the interpretation that most of them are
part of the excess stellar population related to M60/NGC4647. Future 
spectroscopic work will be needed to unambiguously identify the nature of this
$i$-dropout overdensity.

\end{abstract}

\section{Introduction}

  The so-called ``Lyman-break'' technique (Steidel \& Hamilton 1992) is now
widely used to select galaxies at high redshifts. This method relies on
multi-band imaging to identify the characteristic discontinuity --- the
{\it Lyman-break} --- in the spectral energy distributions (SED) of
high-redshift galaxies, which is largely caused by the Lyman-limit
and \Lya\, absorptions of intervening neutral hydrogen along the sightlines to
such galaxies (e.g., Madau 1995). The technique requires imaging in at least
two passbands, one to the blue side of the break and the other to the red side.
Lyman-break makes high-redshift galaxies much fainter in the blue band than in
the red one, or in other words, it makes them seem to ``drop-out'' from the
blue band. For this reason, this method is also known as the ``drop-out''
selection, and the candidates found in this way are usually referred to as 
``dropouts''. 

  As its first application in the $z\approx 6$ regime, Yan, Windhorst \& Cohen
(2003; hereafter YWC03) used this technique in a deep field observed by the
Advanced Camera for Surveys (ACS; Ford et al. 2003) in the default pure parallel
mode (Sparks et al. 2001) soon after its deployment on-board the Hubble Space
Telescope (\hst). In this redshift range, Lyman-break occurs at around 8512\AA\,
in observer's frame (restframe 1216\AA), which is well targeted by the F775W
and F850LP filters through which this deep parallel field was imaged. YWC03
found a large number of F775W-band dropouts, and argued that the vast majority
of them were very likely galaxies at $z\approx 6$. 

  The number density of $z\approx 6$ galaxies inferred from YWC03, however,
seem to be much higher than those in other deep ACS fields (e.g.,
Bouwens et al. 2003; Stanway, Bunker \& McMahon 2003; Dickinson et al. 2004;
Bouwens et al. 2004; Bunker et al. 2004; Yan \& Windhorst 2004, hereafter YW04).
While the cosmic variance could result in significant fluctuation in the number
density (e.g., Somerville et al. 2004; Bouwens et al. 2004), it cannot account
for the large difference between the result of YWC03 and those of others. As 
the true number density of galaxies at $z\approx 6$ is closely related to a 
series of important cosmological questions (e.g., the formation of early 
galaxies; the reionization history of the universe; etc.), it is prudent to 
closely examine the cause of the observed difference. 

   In this paper, we present new results that shed light to the nature
of this overdensity. We first describe the existing and new \hst\, observations
in \S 2 and the data reduction in \S 3. The overdensity of F775W-band dropouts
in YWC03 is scrutinized in \S 4, and a new interpretation is given in \S 5. We 
conclude with a summary in \S 6. For simplicity, we refer to the \hst\, F475W, 
F775W, F850LP, F110W, and F160W passbands as \acsg, \acsi, \acsz, \nicJ, and
\nicH, respectively. The F775W-band dropouts are then referred to as 
$i$-dropouts. The magnitudes are in AB system. All coordinates quoted are in
J2000. The following cosmological parameters from the first-year Wilkinson
Microwave Anisotropy Probe (WMAP) results in Spergal et al. (2003) are
adopted: $(\Omega_M, \Omega_\Lambda, H_0)=(0.27,0.73,71)$. Using the three-year
WMAP results (Spergal et al. 2007) would not change any of our results.

\section{Observations}

  In addition to the ACS parallel field discussed in YWC03 (hereafter referred
to as ``Par1''), we have acquired two sets of new observations to help tackle
the question at hand. To follow-up a subset of the $i$-dropouts found in YWC03,
two regions in Par1 were imaged by the Near Infrared Camera and Multi-Object
Spectrometer (NICMOS) Camera 3 (NIC3). During the course of these NIC3 
observations, the ACS instrument was working in parallel, which resulted in a
new ACS parallel field (hereafter referred to as ``Par2'') that is several
arcminutes away from the original one. The layout of these two ACS fields are
show in Fig. 1. Note that they are in the general direction of the Virgo 
Cluster. We describe all these observations below.

\subsection{ACS parallel observations of Par1}

   The center of Par1 is at $RA=12^h43^m32^s$, $Dec=11^o40'32''$. The ACS data
were taken during \hst\, Cycle 11 soon after the instrument was installed in
March, 2002. In fact, the observations spanned from April 28 to June 19, 2002,
and were acquired during the execution of the Guest Observer (GO) program ID
9043 (PI. Tonry), when it was using the Wide-Field Planetary Camera 2 (WFPC2) as
the primary instrument to image NGC4647, which is a member of the Virgo
Cluster. Only \acsi\, and \acsz\, filters were used for these ACS parallels.
In total, 15 images were taken in \acsi\, and 27 images were taken in \acsz. The
total exposure time in these two bands is 2.65 hours and 4.28 hours,
respectively.

\subsection{NIC3 observations in Par1}

   Our NIC3 observations of Par1 were carried out as a GO program (PID 9780) in
\hst\, Cycle-12. Two regions in Par1, designated as fields ``Sub1'' and
``Sub2'', were observed in $J_{110}$ and $H_{160}$. Due to an unfortunate
system glitch that caused NICMOS enter a short safing-mode period in early
August 2003, our program missed its originally planned visit windows, and its
execution had to stretch from January to December 2004 through rescheduled
visits.

   For each field, the observations were done at 15 dithering positions in
$J_{110}$ and 25 positions in $H_{160}$. The observation mode was MULTIACCUM
with SPARS64 sequence. The exposure time at each dithering position is 512
seconds, resulting in a total integration time of 2.13 hours in $J_{110}$ and
3.56 hours in $H_{160}$, respectively.

\subsection{ACS parallel observations of Par2}
 
   While our NIC3 program was not executed in August 2003 as originally
scheduled, part of its ACS parallel observations were still automatically
carried out on August 5 and 6, 2003 as planned in default.
This resulted in Par2, which is about 8$'$ to the
north of Par1. The field center is $RA=12^h43^m30^s$, $Dec=11^o49'21^s$.
In total, 5 images in $z_{850}$, 12 images in $i_{775}$ and
6 images in $g_{475}$ were taken in this field, giving a net exposure time of
1.00, 1.80 and 0.89 hours in these three bands, respectively.

\section{Data Reduction and Photometry}

   In this section, we discuss the data reduction and photometry of all
relevant \hst\, images.

\subsection{ACS data reduction}

   When the YWC03 paper was written (September 2002), the HST ``On-the-Fly
Reprocessing'' (OTFR) calibration pipeline could only process ACS data up to
the step of geometric distortion correction. The {\it Multidrizzle} package 
(Koekemoer et al. 2002), now a standard tool for mosaicing ACS images using
the drizzle algorithm (Fruchter \& Hook 2002), was then not yet well known
and tested. Therefore, the ACS mosaics of Par1 used in YWC03 were created by
stacking the flat-fielded, distortion-corrected images utilizing the 
well-tested {\it IMCOMBINE} task of IRAF that had been used by the community
for decades.

   For this current study, the ACS data of Par1 were retrieved again from the
\hst\, data archive. In so doing, the requested data were automatically 
reprocessed by the OTFR pipeline using the best reference files (bias, dark 
current, flat field, etc.) currently available. New mosaics were then made from
these reprocessed images by using the now-fully-matured {\it Multidrizzle} 
routine as available in STSDAS under PyRAF. The reduction of the Par2 ACS data
followed the same procedure.

   Some ACS/WFC OTFR-processed images are known to show a quadrant-to-quadrant
``jump'' in background because of the varing residual bias levels in the four
ACS/WFC readout amplifiers (Pavlovsky et al. 2005).
While this bias offset is
minimal in the data of Par2 ($\sim 0.1 e^{-}$ level), it is obvious in most
of the Par1 images ($\sim 10 e^{-}$ level). Although such ``jumps'' do not
have significant impact on the photometry of compact sources in general, they
do affect the sources that are close to quadrant boundaries. Therefore, before
stacking the Par1 data, such jumps were removed by subtracting the background
from each quadrant using our home-grown routines. The quadrant offsets were
reduced to $\sim 1 e^{-}$ level after this process.

    The final mosaics were created following the standard drizzling procedures
implemented in {\it Multidrizzle}, and were normalized to unit exposure time
(count per second). A square kernel was used in drizzling, and
the linear size of drop was set to 0.9 (the ``final\_pixfrac'' parameter of 
{\it Multidrizzle}). To get a finer spatial resolution, the drizzle scale (the
``final\_scale'' parameter) was set to 0.6, which resulted in an output pixel
scale of 0.03$''$/pixel. For each field, the output mosaics in different
passbands were all registered to the same reference position in the process.
The absolute astrometry of the final mosaics was calibrated by using the
compact objects visible in the Digital Sky Survey images.

    In order to provide a reference frame for Par1 NIC3 data reduction, another
version of Par1 ACS mosaics was also created by setting drizzle scale to 1.0,
i.e., preserving the native pixel scale (0.05$''$/pixel) of the ACS/WFC. We
will refer to these lower-resolution mosaics and the higher-resolution ones
mentioned above as ``LOWRS'' and ``HIGHRS'' ACS mosaics, respectively. They
each serve different purposes in the analysis process. 

    As the original pixels were resampled during drizzling (and during
geometric distortion correction as well), the final mosaics all have correlated
pixel noise. The signal-to-noise ratio (S/N) of a given object derived directly
from the final mosaic would therefore be overestimated, and hence its
photometric error would be underestimated. We used a routine, kindly provided
to one of us (HY) by Dr. Mark Dickinson, to calculate the correlation amplitude
and hence to calculate proper statistical error associated with each pixel. The
error map (so-called ``RMS'' map) derived in this way are widely used,
e.g., in the data analysis of GOODS (e.g., Dickinson et al. 2004).

\subsection{NIC3 data reduction}

    All data obtained by NICMOS show a number of persistent anomalies that
cannot currently be handled by the OTFR pipeline. Therefore, the OTFR-processed
NIC3 data of Sub1 and Sub2 fetched from the archive were further reduced before
stacking. 

   An additional count-rate and wavelength dependent non-linearity of
NICMOS was recently identified by the STScI NICMOS team (NICMOS Instrument
Science Reports 2006-001, 002 and 003; see also
http://www.stsci.edu/hst/nicmos/performance
/anomalies/nonlinearity.html).
This non-linearity was removed from the data by using the routine provided
by the team.

   The quadrant bias of NICMOS also has a stochastic behavior similar to
what described above for ACS. As a result, a bias-corrected NICMOS image using
the standard bias reference file usually show ``pedestals'' from quadrant to
quadrant. Such pedestals were removed by using the {\it PESKY} task included
in the STSDAS package. 

   All OTFR-processed NIC3 data have a persistent, additive pattern across
entire field. This pattern is different in different passbands, but is rather
stable with respect to time. To remove this pattern, a ``pattern image'' in
each band was first created by stacking all the pedestal-corrected images in
this band. The pattern image was then subtracted from each individual image.

   After all the above steps, the individual images were combined using the
series of drizzle tasks included in the {\it DITHER} package of STSDAS in
IRAF. Each input science image was weighted by a weight image calculated from
its associated error array stored in the [ERR,2] extension of the 
OTFR-processed image. For the purpose of matched-aperture photometry that will
be discussed below, the LOWRS Par1 ACS $z_{850}$-band mosaic was binned in
$2\times 2$ and then used as the reference. The NIC3 images were all mapped
(i.e., rotated and registered)
to this binned ACS image during the drizzling process. The drizzle 
scale was set to 0.5, which resulted in a final resolution of 0.10$''$/pixel,
i.e., the same as the binned LOWRS ACS image. The final mosaics were
normalized to unit exposure time as well.

   The associated RMS maps of all these NIC3 mosaics were derived in the way
similar to what described above for the ACS mosaics.

\subsection{ACS source detection}

   As compared to the ACS mosaics used in YWC03, the new mosaics created
through the imporved reduction process (see \S 3.1) are of much higher quality,
and hence we can push to a significantly lower thresold for source detection.

   SExtractor (Bertin \& Arnouts 1996) was run on the HIGHRS mosaics in 
dual-image mode to extract source catalogs for both Par1 and Par2. 
``MAG\_AUTO'' magnitudes were used to best approach the total magnitudes.
A number of source catalogs were
obtained for different purposes. The \acsi-band and \acsz-band mosaics
were used alternatively as the detection images to perform ``matched-aperture 
photometry'', i.e., measuring the magnitudes of a given source in different
bands through the same aperture defined by its appearance on the detection
image. We will refer to these catalogs as ``$i_{775}$-based'' and 
``$z_{850}$-based'' ACS catalogs, respectively. Source detection was performed
at the threshold of 0.8~$\sigma$, using a $5\times 5$ Gaussian filter with a 
full-width-at-half-maximum (FWHM) of 2.5 pixels. A minimum number of 4 
connected pixels were required for a source to be included. 

   As Par1 was observed in parallel mode during the very early stage of ACS,
the data were all taken at a high gain setting of 4.0~$e^-$/ADU, probably for
safety reason (i.e., avoiding saturation). In order to make the zeropoints
derived at a gain setting of 1.0~$e^-$/ADU universally applicable, the ACS OTFR
pipeline multiplies the flat-field reference file by the gain of the object
image before flat-field correction. In the end, an OTFR-processed ACS image
always has an effective gain of 1.0~$e^-$/ADU. When YWC03 was written,
unfortunately, the authors were not aware of this --- at that time 
poorly-documented --- step in the OTFR processing,
and made a redundant correction for the gain by adding
$2.5\times log10(4.0)=1.505$ to the zeropoints. As a result, while the
candidate selection in YWC03 remains valid, all magnitudes reported in that
paper were too faint by 1.505 mag. This error has been reported in 
a footnote in YW04.

   In this current study, the zeropoints published in the latest ACS 
handbook were used; namely, 26.068, 25.654 and 24.862 mag for $g_{475}$, 
$i_{775}$ and $z_{850}$, respectively. The 3~$\sigma$ limits within a circular
aperture of 0.1$^{''}$ in radius are 29.1 and 28.3 mag in \acsi\, and \acsz\,
in Par1, and 28.9, 28.7 and 27.9 mag in \acsg, \acsi\, and \acsz\, in Par2,
respectively. 

\subsection{NIC3 source detection}

   Photometry of Sub1 and Sub2 NIC3 mosaics in Par1 was also done in
dual-image mode, using $J_{110}$ and $H_{160}$ mosaics alternatively as the
detection images.
Similarly, we will refer to these catalogs as ``\nicJ-based'' and 
``\nicH-based'' catalogs, respectively. The detection filter was a 
$5\times 5$ Gaussian filter with a FWHM of 3 pixels, the detection threshold
was set to 0.4~$\sigma$, and a minimum of 2 connecting pixels were required.
Again, MAG\_AUTO magnitudes were adopted.
As the magnitude zeropoints
provided by the OTFR pipeline were no longer valid after the non-linearity
correction described in \S 3.2, we followed the recipe given by the NICMOS
team to derive the zeropoints as 23.242 and 23.139 mag in $J_{110}$ and
$H_{160}$, respectively.

   During this process, ACS magnitudes of the NIC3 sources were also extracted
in order to properly measure their ACS-to-NICMOS colors for later analysis.
As the NIC3 mosaics were created by registering to the reference frame defined
by the $2\times 2$ binned LOWRS ACS mosaics, matched-aperture photometry between
NIC3 and ACS passbands was fairly straightforward. The detection images were
still the NIC3 mosaics, and the object images were $2\times 2$ binned ACS 
mosaics that were convolved by the PSF of the \nicJ\, images (derived using the
TinyTim tool). The binning operation was sum, which preserves the original ACS
magnitude zeropoints. 

\section{Overdensity of \boldmath $i$-dropouts}

   Using the newly produced ACS mosaics of Par1, new NIC3 data obtained at two 
pointings within Par1, and a new ACS parallel field Par2 for comparison, here
we will re-examine the overdensity of $i$-dropouts in Par1 reported by YWC03.

\subsection{Overdensity of \boldmath $i$-dropouts in Par1}

   YWC03 reported 30 $i$-dropouts in Par1, whose \acsz\, magnitudes range from
26.8 to 28.3 mag. Using the correct zeropoints, their true $z_{850}$-band 
magnitude range should be 25.3 to 26.8 mag (see \S 3.3). The
corresponding cumulative surface density thus is $\sim$ 2.7 per arcmin$^2$
to $z_{850}=26.8$ mag. Taken at its face value, this number is 3--4 times
higher than those in other deep, high galactic-latitude ACS fields, including
the Hubble Ultra Deep Field (HUDF; Beckwith et al. 2006).
The $i$-dropouts in YWC03 were selected based on their invisibility in the
\acsi\, image, i.e., by selecting sources that have $S/N\geq 5$ in
$z_{850}$ but have $S/N<2$ in $i_{775}$. As a result, about 2/3 of those
objects have $i_{775}-z_{850}>2.0$ mag. Other studies utilizing ACS data 
(including YW04) all have adopted less stringent selection criterion of 
$i_{775}-z_{850}\geq 1.3$ mag or $i_{775}-z_{850}\geq 1.5$ mag. 
Therefore, the high surface density of YWC03 is even more difficult to be 
reconciled with all the other later work.

   To study this problem in detail, here we reselect $i$-dropouts in Par1
from the new $z_{850}$-based ACS catalog as described in \S 3.3. We only 
consider sources with $S/N\geq 5$ in \acsz, and adopt the color criterion of
$i_{775}-z_{850}\geq 1.3$ mag for ease of comparison with other results. The
candidates thus selected are visually examined to eliminate false detections
such as residual cosmic rays and image defects. The final $i$-dropout sample 
consists of 356 objects to $z_{850}=28.0$ mag. This is much deeper than what
we achived in YWC03, and is the result of the much improved data quality. In
particular, 198 of them have $z_{850}\leq 26.8$ mag (among which 21 have
$i_{775}-z_{850}\geq 2.0$ mag), which is a factor of 20 more than the 
$i$-dropouts selected in the HUDF
(YW04; Bunker et al. 2004) to the same depth. This confirms -- and reinforces
-- the unusually large overdensity of $i$-dropouts seen by YWC03. The
distribution of these $i$-dropouts in Par1 are shown in Fig. 2.

   A necessary condition for an $i$-dropout being a legitimate candidate of
galaxy at $z\approx 6$ is that it must be a dropout in bluer bands. 
However, Par1 does not have any data to the blue of \acsi. 
Fortunately, as a subset of
these $i$-dropouts have NIC3 observations, we can use their ACS-to-NIC3 colors
to judge whether they are consistent with being galaxies at $z\approx 6$.

\subsection{Optical-to-IR colors of \boldmath $i$-dropouts in Par1}

   In total, 21 $i$-dropouts are detected in our NIC3 data. Fig. 3 displays 
the image cut-outs of one such objects. The optical-to-IR color-color diagrams
of all these sources are shown in the left panel of Fig. 4.
For comparison, the right panel of Fig. 4 shows the similar color-color
diagrams from the HUDF.

   It is clear that the $i$-dropouts in Par1 reside in the similar regions in
the optical-to-IR color space as the $i$-dropouts in the HUDF. In particular,
they are well separated from brown dwarf stars and E/S0 galaxies at
lower redshifts, which are the most common contaminators to $z\approx 6$
galaxy samples selected by the drop-out technique.

\subsection{A similar overdensity of \boldmath $i$-dropouts in Par2}

   In order to compare with the results obtained in Par1, we perform
$i$-dropout selection in Par2 in a similar way. As this field has $g_{475}$
data, an additional constraint of non-detection in $g_{475}$ at 2~$\sigma$
level is imposed for the selection. The final $i$-dropout sample consists of
20 objects to $z_{850}=26.5$ mag, which, while much less than the ones in Par1,
is still a factor of $\sim 7$ more than those found in the HUDF to the same
depth; in other words, the overdensity of $i$-dropouts also seems to present in
Par2.

\section{Alternative Interpretation of the Overdensity}

  The analysis in \S 4 indicates that the $i$-dropouts in Par1 are consistent
with being galaxies at $z\approx 6$, and that the overdensity is also present
in Par2. If this interpretation is true, we will have to draw the conclusion
that the overdensity of $i$-dropouts in this region is due to an extremely
unusual concentration of $z\approx 6$ galaxies over $\sim 10$~Mpc scale. In this
section, however, we present another possibility.

\subsection{Excess of field population in Par1 \& Par2}

   The alternative interpretation is triggered by an unusual field population
seen in both Par1 and Par2. This is illustrated in Fig. 5, which shows the
 \acsz\, vs. \acsi$-$\acsz\, color-magnitude diagrams in these fields using the
\acsz-based catalogs, together with the corresponding diagrams constructed in
the HUDF based on the photometry from YW04. The left panel only plots the
sources at $S/N>5$, while the right panel includes all the sources at $S/N>3$.
A striking feature in this figure is the large concentration of objects at the
faint end in the two parallel fields, which is not seen in the HUDF (and other
deep ACS fields). Due to the shallower depth in Par2, this population is at
the verge of being detectable at $S/N>5$; however, it gets very prominent if
the threshold is lowered to $S/N>3$. 
YWC03 noticed only a hint of this excess in Par1; the
source detection back then did not reach a sufficient depth that could 
unambiguously confirm its existence. 

   This excess of faint field objects in Par1 and Par2 is present in all 
passbands. To quantify this excess, Fig. 6 compares the surface densities of
the detected sources in these two fields and the ones in the HUDF, derived
at two different threshold levels of $S/N>3$ (left panel) and $S/N>5$ (right
panel). Again, while the excess in Par2 is only detectable when the sources
of low $S/N$ are included, the excess in Par1 is obvious even at a high
source detection threshold. This excess starts at about 26.0 mag in both Par1
and Par2, and extends to a much fainter level. In particular, the
deep mosaics in Par1 allow this excess to be traced to at least $\sim 28.2$ and
27.2 mag in \acsi\, and \acsz, respectively. To further ensure the existence of
the excess, we split the available images of Par1 into two groups and
created shallower stacks in both bands. The same excess was detected as well.

   The $i$-dropout surface densities are also superposed in Fig. 6. We point
out that the $i$-dropout excess in the two parallel fields resides in the 
region where the field population excess happens. This is further discussed
later in the paper.

\subsection {Resolving diffuse light from interacting galaxy pair}

  What is the nature of this field population excess? The answer probably
lies in the proximity of Par1 and Par2 to the galaxy pair M60/NGC4647, which
is several arcminutes away to the South (see Fig. 1). Extended diffuse
emissions of different natures have been detected around some nearby galaxies,
for example in NGC 5907 (Shang et al. 1998; Zheng et al. 1999) and in M31
(e.g., Ferguson et al. 2002). While there is no previous report in the 
literature, we speculate that such extended diffuse emission also exists around
M60/NGC4647, and propose that the excess population is due to the fact that we
have resolved the diffuse light extending from this interacting pair. 
If we adopt an average distance of $\sim 15.7$~Mpc
to M60/NGC4647 (from NASA/IPAC Extragalactic Database), Par1 and Par2 are 
$\sim$ 32 and 69~kpc away from the center of pair, respectively. Therefore, 
what we have detected are most likely the halo stars of the two galaxies and/or
the tidal ``debris'' repelled from this pair during the course of the 
interaction.

  Here we exam the property of this excess more closely. We only use Par1 for
this purpose, as the data in this field are much deeper than in Par2. Fig. 7
plots the surface densities of $S/N>5$ sources in Par1 as functions of 
magnitudes in \acsi\, (left) and \acsz\,(right), and compares them with the 
results in the HUDF. The insets show the net excess after subtracting the 
surface density calculated in the HUDF. Tab. 1 lists the surface densities
of the net excess as a function of magnitudes. 

   Taken at face value, the net excess in Par1 as shown in Fig. 7 has a total
magnitude of 15.0 and 15.7 mag in \acsi\,
and \acsz, respectively. The average surface brightness from the net excess
(i.e., spreading the total integrated light of the net excess over the entire
field of Par1), on the other hand, is 26.6 and 27.2 mag/arcsec$^2$ in \acsi, 
and \acsz, respectively.  If this excess is
isotropic with respect to M60/NGC4647, the total magnitude integrated over
an annulus 3.4$^{'}$ in width (i.e., about the width of one ACS field) at the
distance of Par1 (centering on the galaxy
pair) is 12.2 mag in \acsi\, and 12.9 mag in \acsz. For comparison, the total
flux of M60 in Johnson I-band is 8.46 mag (or 0.925~Jy, i.e., 8.98 mag in AB)
within 84$^{''}$ aperture (Boroson, Strom \& Strom 1983). Therefore, the total
diffuse light within this annulus is about a few percent of the central galaxies.

The net excess peaks at about 27.1 and 26.6 mag in \acsi\, and \acsz.
If these objects are indeed at the distance of Virgo Cluster, these values
correspond to absolute magnitudes of $-3.9$ and $-4.4$ mag, respectively,
which are in the regime of giants/supergiants or low luminosity globular 
clusters.

   We shall point out that the excess that we detected relates more closely to
M60/NGC4546 rather than to the Virgo Cluster in general. A supporting evidence
comes from the spatial gradient of this excess, which decreases in amplitude
when moving away from the pair. This is demonstrated in Fig. 8, which plots the
spatial distribution of all $S/N>5$ sources at \acsz$>26.0$~mag detected in the
\acsz-based catalog of Par1. The gradient is obvious: there are more sources to
the south, which is the direction where the galaxy pair lies. 

   Another piece of evidence
is that the amplitude of this excess (see Fig. 7 and Tab. 1) cannot be 
attributed to the intracluster stellar populations of the Virgo Cluster alone. 
We only consider Virgo's intracluster globular clusters (IGC) and red giants
(IRG), as they are the only two type of stellar objects that are within the
luminosity range of the detected excess. The {\it total} contribution from
Virgo's intracluster globular clusters (IGC) at $> 26.0$~mag, calculated using
the Virgo's IGC surface density at the bright-end (e.g., Williams et al. 2007a;
Cohen et al. 2003) and the symmetry of globular cluster luminosity function, is
only $\sim 0.3$--3.0~arcmin$^{-2}$ (for any reasonable color transformation
from F814W to F775W), depending on the proximity to a member galaxy. On the
other hand, the deep \hst\, surveys of Virgo's IRG (e.g., Ferguson, Tanvir \&
von Hippel 1998; Durrell et al.  2002; Williams et al. 2007b) show that their
cumulative surface density to $\sim 28.0$ mag can be as high as 
$\sim 110$~arcmin$^{-2}$ (again for any reasonable F814W to F775W 
transformation). However, this is still a factor of 5--6 lower than
what shown in Fig. 7 and Tab. 1.

\subsection{Relation between $i$-dropout overdensity and field object overdensity}

   We suggest that the $i$-dropout overdensity observed in Par1 \& Par2
might be related to the field object overdensity in the same regions.
The field object excess becomes significant at 26.0 mag in both bands, and
extends to at least 28.2 mag in \acsi\, and 27.2 mag in \acsz\, in the deeper
Par1 images (see the insets in Fig. 8) where the excess can be traced to a 
faint level at high confidence. The $i$-dropouts selected in these fields
resides in the \acsz\, magnitude range that the field object excess spans
(Fig. 6). The spatial distribution of the $i$-dropouts also seems to follow
the gradient that seen in the distribution of the excess field population
(see Fig. 2 \& 7).

   This excess field population is also red in color. Fig. 9 shows the surface
densities of the sources in the above mentioned magnitude ranges as a function
of \acsi$-$\acsz\, color. For comparison, the surface densities of the HUDF 
sources in the same magnitude ranges are also shown. The insets plot the net
excess using the HUDF result as the reference. The excess population peaks at
\acsi$-$\acsz$\approx 0.35$ mag in the \acsi\, image, and peaks at 
\acsi$-$\acsz$\approx 0.60$ mag in the \acsz\, image. Note that the distribution
extends continuously beyond \acsi$-$\acsz$\geq 1.3$ mag, which is the regime
where the $i$-dropouts are selected.

   If the $i$-dropouts are indeed some stellar objects related to 
M60/NGC4647, what type of stars could they be? Thus far we have
not yet been able to identify the satisfactory candidates. As Fig. 4 shows,
brown dwarfs are not likely to contribute much to our $i$-dropout sample 
because their optical-to-IR colors are different. In addition, their
luminosities are too low. The only possibility then left is dusty stellar
population, for example, Asymptotic Giant Branch (AGB) stars. However, the
optical-to-IR colors of AGB stars are quite different from the $i$-dropouts as
well. This is illustrated in Fig. 10, where the synthesized colors of
AGB stars generated from the model spectra of Lancon \& Mouhcine (2002)
are superposed on the color-color diagrams of the $i$-dropouts in Par1.
We should also point out that no external dust extinction to any known types
of stars could produce such colors; no know extinction law could create such
a red color in $i_{775}-z_{850}$ and yet keep $z_{850}-J_{110}$ and
$J_{110}-H_{160}$ colors around zero.

   The red, excess field objects that we have discovered here seem to have some
connection to the so-called ``Red Halo Phenomenon'' (Zackrisson et al. 2006),
which referrs to the very red colors of the halos found around some galaxies
(e.g., Zibetti, White \& Brinkmann 2004; Zibetti \& Ferguson 2004; Taylor et al.
2005). Such red halos cannot be easily explained by standard stellar 
populations. The $i_{775}-z_{850}$ color shown in Fig. 9 quantitively agrees
with that of the halo around an edge-on disk galaxy in the HUDF reported by 
Zibetti \& Ferguson (2004). Therefore, the excess field objects that we see
might represent the first example where such red halos have been resolved, and
the $i$-dropouts might be the most extreme sources among this population that
is most difficult to explain.

\section{Summary}

   In this paper, we use new and existing \hst\, data to closely examine the
large overdensity of $i$-dropouts that YWC03 found in a deep ACS parallel 
field. The new data include NIC3 imaging of a subset of the $i$-dropouts, and
a set of new ACS parallel data in a nearby region obtained during the NIC3 
observations. We confirm that a factor of $\sim 20$ overdensity of $i$-dropouts
does exist in the parallel field of YWC03, and that the optical-to-IR colors of
these objects are similar to those selected in other deep ACS fields such as
the HUDF. 

   However, these $i$-dropouts might {\it not} be galaxies at $z\approx 6$ as
one would expect. During the course of our investigation, we have found a large
excess of discrete sources in both the field of YWC03 and the new ACS parallel
field. This excess is most likely due to a stellar population that is related
to M60/NGC4647, an interacting galaxy pair in the Virgo Cluster, which lies
several arcminutes to the south of the ACS parallel fields. At the distance
of the Virgo Cluster, the ACS parallel fields are several tens of kpc away
from the galaxy pair, and hence the stellar objects are likely tidal ``debris''
expelled from the pair and/or distant halo stars belong to these two galaxies.
Both the spatial distribution and the amplitude of the excess indicate that
it is a population related more closely to the galaxy pair rather than to the
Virgo's intracluster populations.

   The $i$-dropouts in the field of YWC03 are within the magnitude range where
the field object excess occurs. Their distribution also shows similar spatial
gradient of the field object excess. Most importantly, the field object excess
is red in color, and the red wing of its color distribution extends well into
the regime where the $i$-dropouts were selected. For these reasons, we believe
these $i$-dropouts are part of the stellar objects that constitute the field
object excess rather than galaxies at $z\approx 6$. 

   Nevertheless, we should caution that we still cannot completely rule out
the possibility that these $i$-dropouts might indeed be $z\approx 6$ galaxies.
On one hand, we do not have any definite evidence against this interpretation.
On the other hand, we have not yet found a stellar population that can
reasonably explain the colors of these $i$-dropouts. To unambiguously identify
their nature, spectroscopic observations of these objects are in demand. As the
brightest sources in our sample are brighter than 26.0 mag in \acsz, this 
might be feasible with the existing spectrographs at our current 8--10m class
telescopes.

\acknowledgments

    The authors thank the referee for the useful comments. We also thank Seth
Cohen, Inese Ivans, Daisuke Kawata,  Huub R\"{o}ttgering and Fran\c{c}ois 
Schweizer for helpful discussions.
Support for program HST-GO-09780.* was provided by NASA through a grant from
the Space Telescope Science Institute, which is operated by the Association of 
Universities for Research in Astronomy, Inc., under NASA contract NAS 5-26555.
This research has made use of the NASA/IPAC Extragalactic Database (NED) which
is operated by the Jet Propulsion Laboratory, California Institute of 
Technology, under contract with the National Aeronautics and Space 
Administration.

\clearpage

\begin{deluxetable}{ccc}
\tablecaption{Surface density of the $S/N>5$ source net excess in Par1 as a function of magnitude}
\tablewidth{0pt}
\tablehead{
\colhead{magnitude} &
\colhead{$\Sigma$(\acsi) (arcmin$^{-2})$} &
\colhead{$\Sigma$(\acsz) (arcmin$^{-2})$}
}
\startdata

25.55  &  0.51 &   1.35 \\
25.65  &  1.96 &   0.93 \\
25.75  &  2.52 &   2.62 \\
25.85  &  5.13 &   2.51 \\
25.95  &  5.54 &   4.67 \\
26.05  &  7.11 &   6.99 \\
26.15  &  9.60 &  13.13 \\
26.25  & 14.58 &  15.86 \\
26.35  & 18.93 &  19.76 \\
26.45  & 29.74 &  29.16 \\
26.55  & 34.35 &  30.55 \\
26.65  & 36.93 &  24.61 \\
26.75  & 38.54 &  18.89 \\
26.85  & 42.51 &  12.57 \\
26.95  & 43.04 &   8.78 \\
27.05  & 47.17 &   7.04 \\
27.15  & 45.84 &   8.32 \\
27.25  & 43.59 &   2.24 \\
27.35  & 36.64 &   --- \\
27.45  & 39.49 &   --- \\
27.55  & 37.16 &   --- \\
27.65  & 26.10 &   --- \\
27.75  & 24.45 &   --- \\
27.85  & 18.44 &   --- \\
27.95  & 11.06 &   --- \\
28.05  &  9.66 &   --- \\
28.15  &  0.58 &   --- \\

\enddata

\tablenotetext{1.} {The net excess is as shown in the insets in Fig. 7, which is
calculated using the count in the HUDF as the reference.}

\end{deluxetable}

\clearpage
\begin{figure}
\plotone{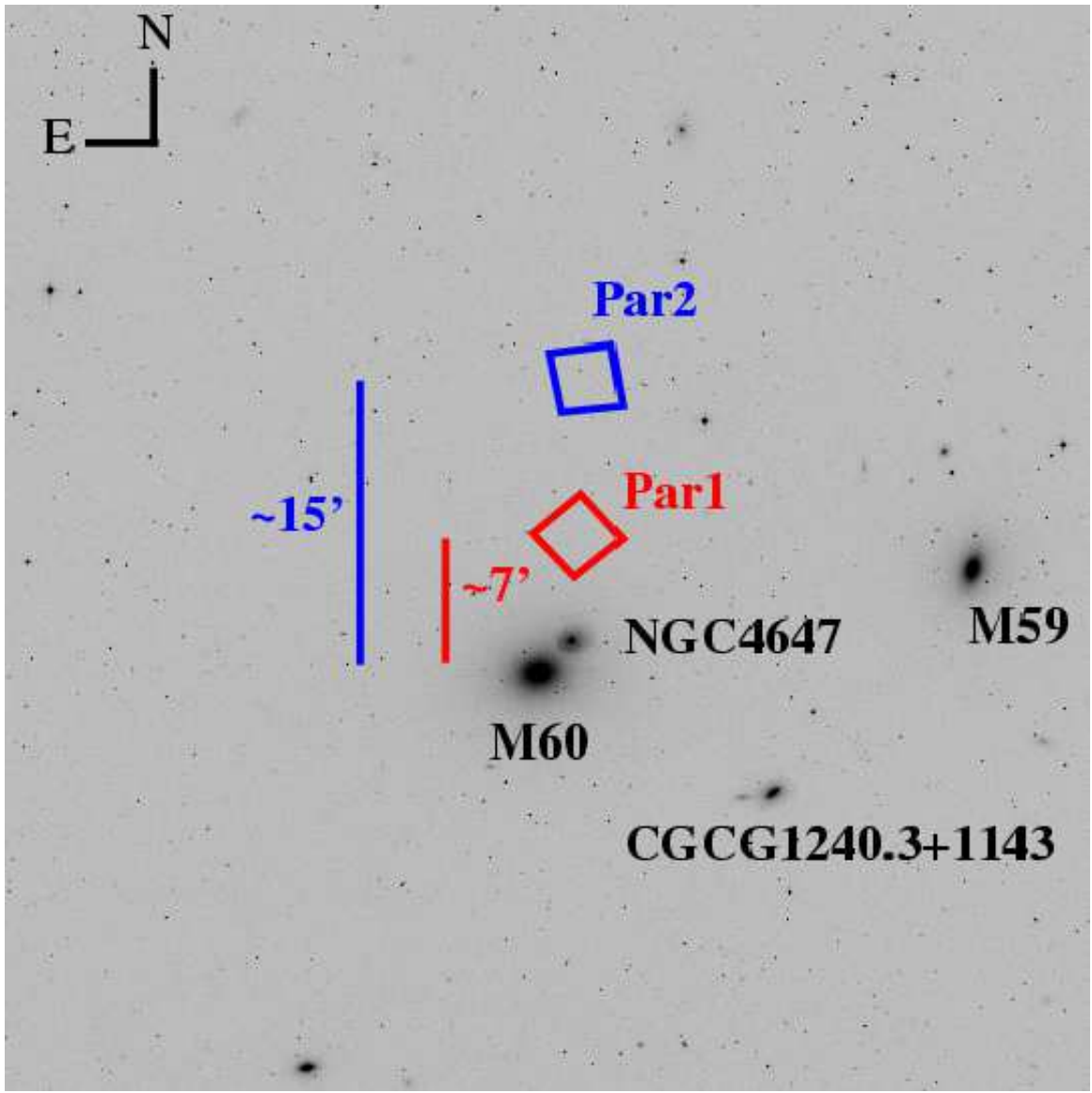}
\caption{The locations of the fields studied in this work are shown on top of
the image from the Digital Sky Survey. Both
fields are several arcminutes away from the interacting galaxy pair, 
M60 \& NGC4647, which are members of the Virgo Cluster. Par1 was taken in
parallel mode by the ACS when the WFPC2 was observing NGC4647, while Par2 was
taken by the ACS in parallel when NICMOS was observing two regions in Par1.
}
\end{figure}

\clearpage
\begin{figure}
\plotone{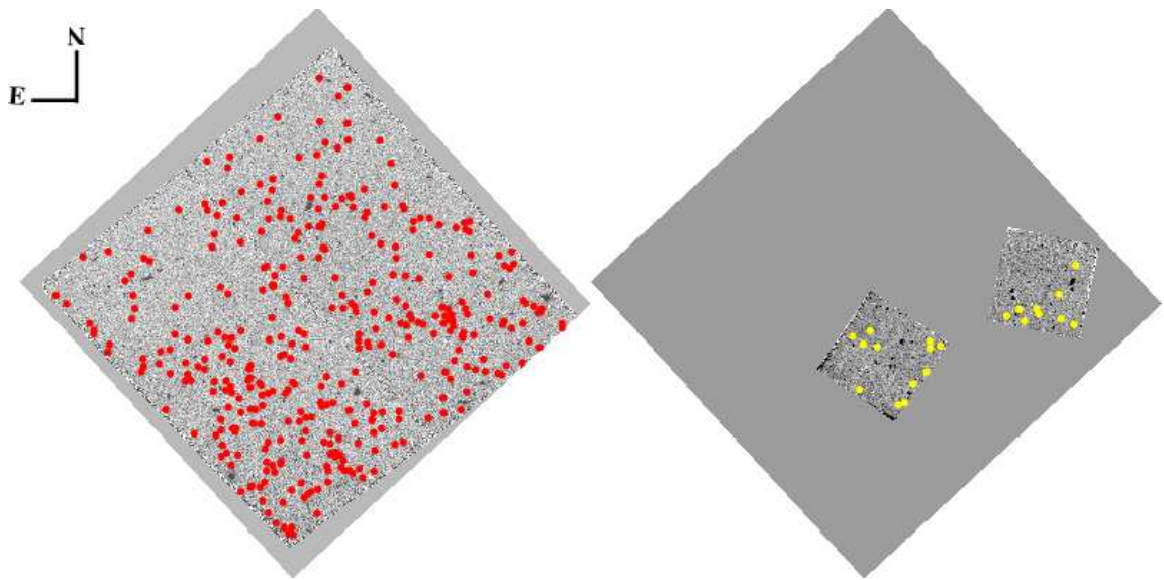}
\caption{The left panel shows the locations of the $i$-dropouts
in Par1 as red dots on top of the \acsz-band mosaic of this field. The yellow
dots in the right panel show the locations of the $i$-dropouts that are
detected the NIC3 mosaic, which is the image in the background (note that
the objects that are seemingly close to the edge are actually well within the
region of good coverage).
}
\end{figure}

\clearpage
\begin{figure}
\plotone{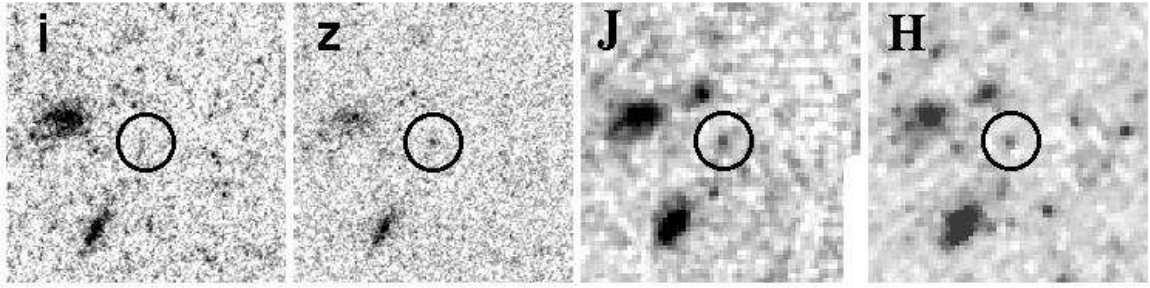}
\caption{The image stamps of one $i$-dropout in our sample. The stamps are
6$^{''}$ on a side, and the circles in the middle are 0.2$^{''}$ in radius.
}
\end{figure}

\clearpage
\begin{figure}
\plotone{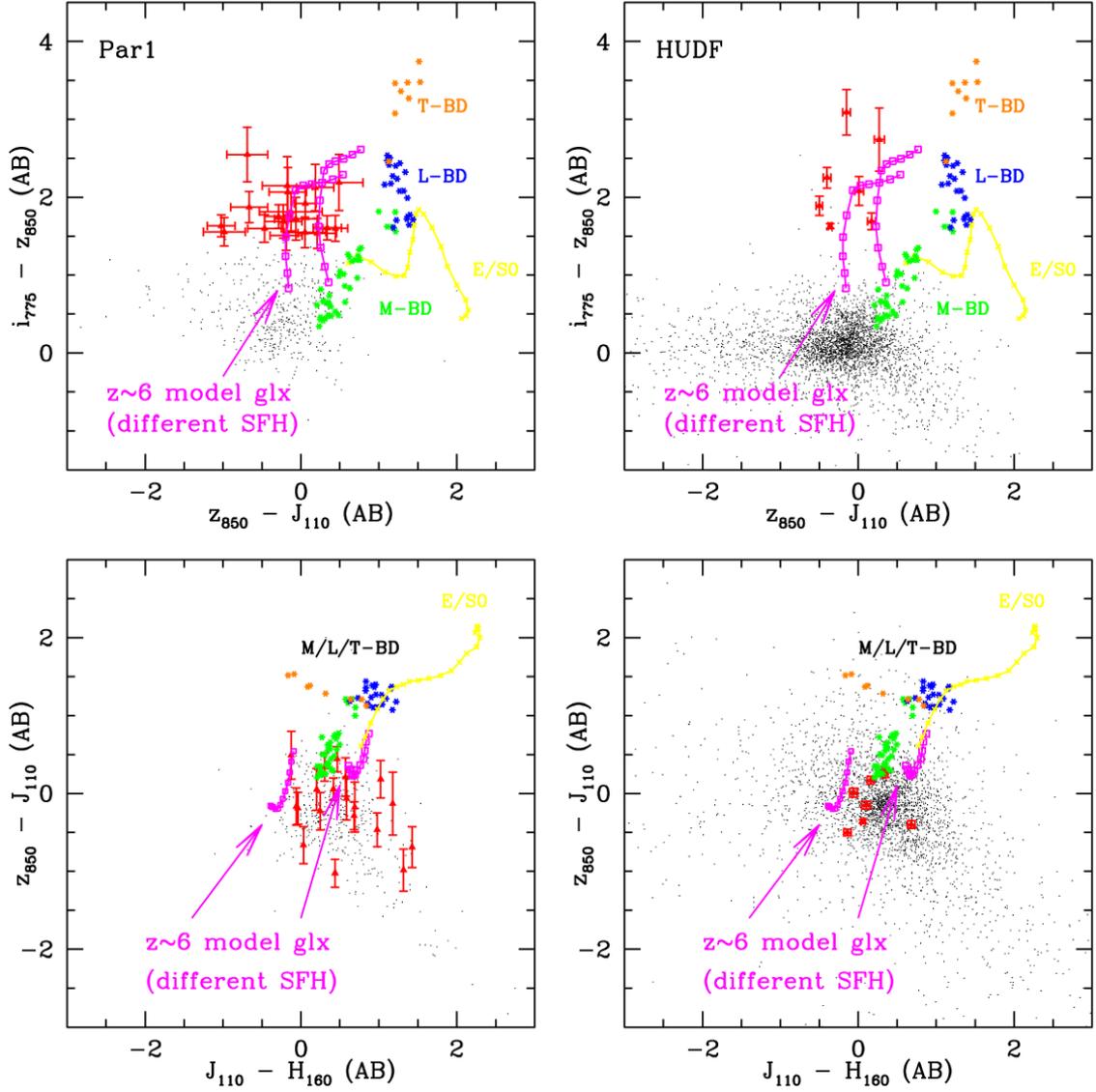}
\caption{The optical-to-IR colors of the $i$-dropouts in Par1 (left) are
compared with those of the $i$-dropouts in the HUDF (right). The $i$-dropouts
are the red points with error bars, while the field objects are shown as the
black dots. M-, L- and T-type brown dwarf stars are the green, blue and brown
asterisks, respectively. The colors of E/S0 galaxy at $z=1$--3 are shown
as the yellow crosses. Two model galaxies at $5.5\leq z\leq 6.5$, constructed
using the models of Bruzual \& Charlot (2003) of different star formation
histories, are shown as the open magenta squares. The $i$-dropouts in Par1
occupy a similar region in the color space as do those in the HUDF.
}
\end{figure}

\clearpage
\begin{figure}
\plotone{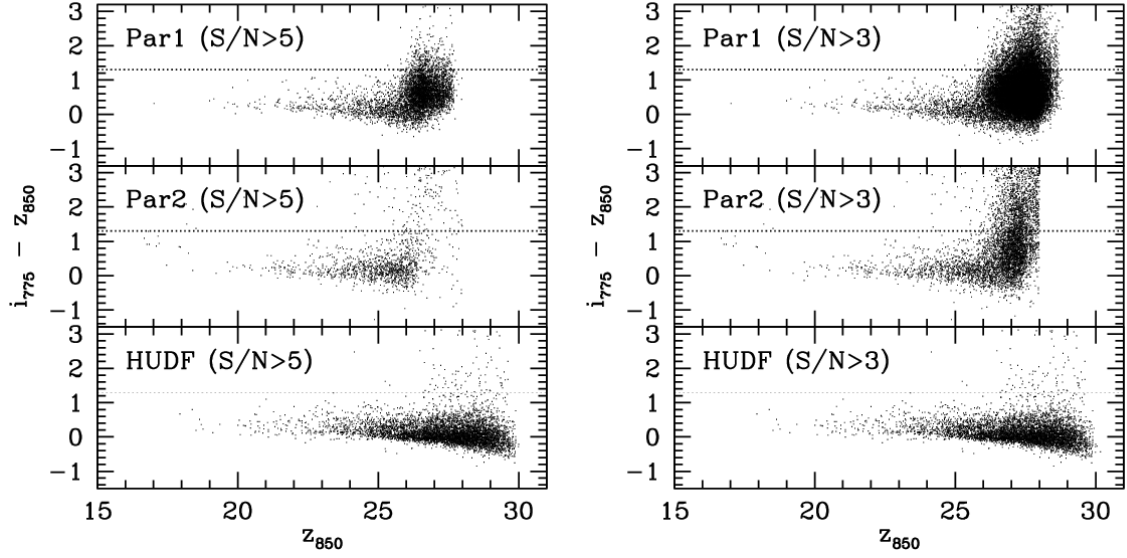}
\caption{Comparison of the \acsz\, vs. \acsi$-$\acsz\, diagrams between the
parallel fields and the HUDF. The objects plotted in the left panel are
limited to the sources with $S/N>5$, while those in the right panel are limited
to the sources with $S/N>3$. The dotted lines show the \acsi$-$\acsz$=1.3$
color threshold above which the $i$-dropouts are selected. A unique feature
in the two parallel fields is the large excess objects at the faint end, which
is not seen in the HUDF. Due to its shallower depth, Par2 does not show this
feature very clearly if only the sources at $S/N>5$ are included; the excess
becomes prominent when sources to $S/N=3$ are also included.
}
\end{figure}

\clearpage
\begin{figure}
\plotone{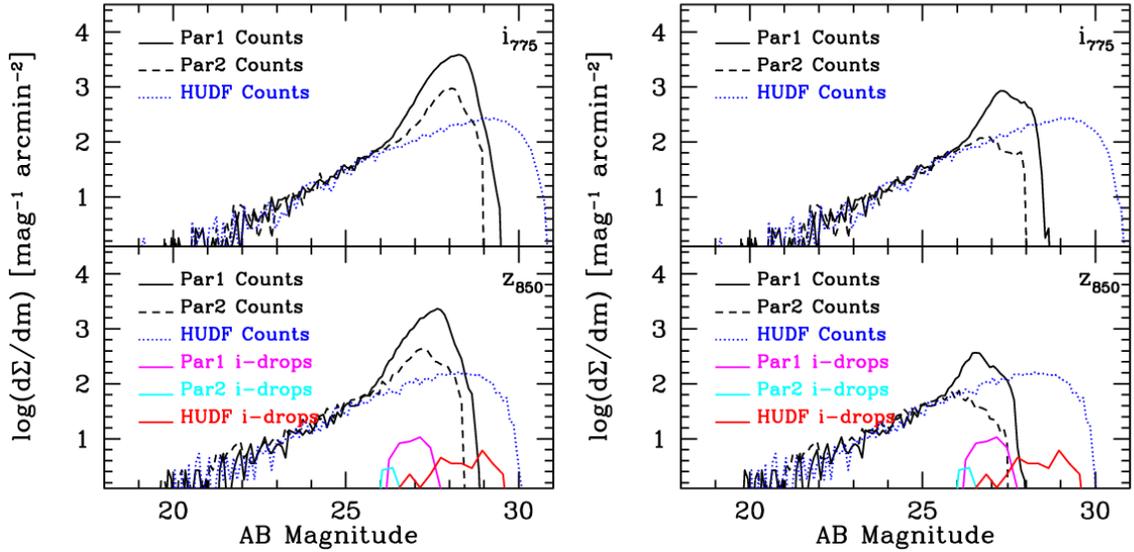}
\caption{Comparison of the field source surface densities in the two parallel
fields (solid and dashed black curves) and in the HUDF (blue dotted curves) at
two threshold levels.
The \acsi-based catalogs are used for the statistics in \acsi-band (top), while
the \acsz-based catalogs are used for the statistics in \acsz-band (bottom). 
The left panel shows the statistics based on sources with $S/N>3$, while the
right panel shows the statistics based on sources with $S/N>5$. The object
excess in the two parallel fields is obvious. The surface densities of 
$i$-dropouts (all have $S/N>5$ in \acsz) in Par1, Par2 and HUDF are also shown
as the magenta, cyan and red curves, respectively.
}
\end{figure}

\clearpage
\begin{figure}
\plotone{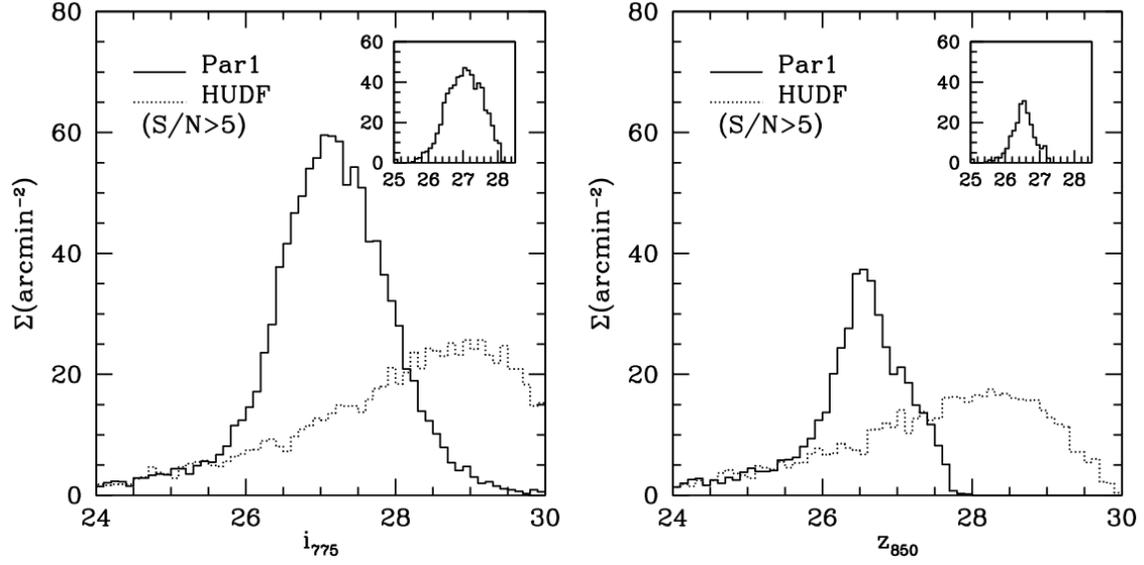}
\caption{Field source excess of Par1 as a function of magnitude. The solid
histograms are the surface densities in Par1, and the dotted histograms are
those in the HUDF. The \acsi-based catalogs are used for the figure in the
left panel, while the \acsz-based catalogs are used for the one in the right.
Only sources at $S/N>5$ are included for the statistics. The insets show the
net excess using the HUDF as the reference. The excess extends to $\sim$ 28.2
and 27.2 mag in \acsi\, and \acsz, respectively.  The average surface 
brightness of the net excess corresponds to 26.6 and 27.2 mag/arcsec$^2$ in
\acsi, and \acsz\, respectively. The total magnitudes of the net excess 
integrated over Par1 are \acsi$=$15.0 and \acsz$=$15.7 mag. 
}
\end{figure}

\clearpage
\begin{figure}
\plotone{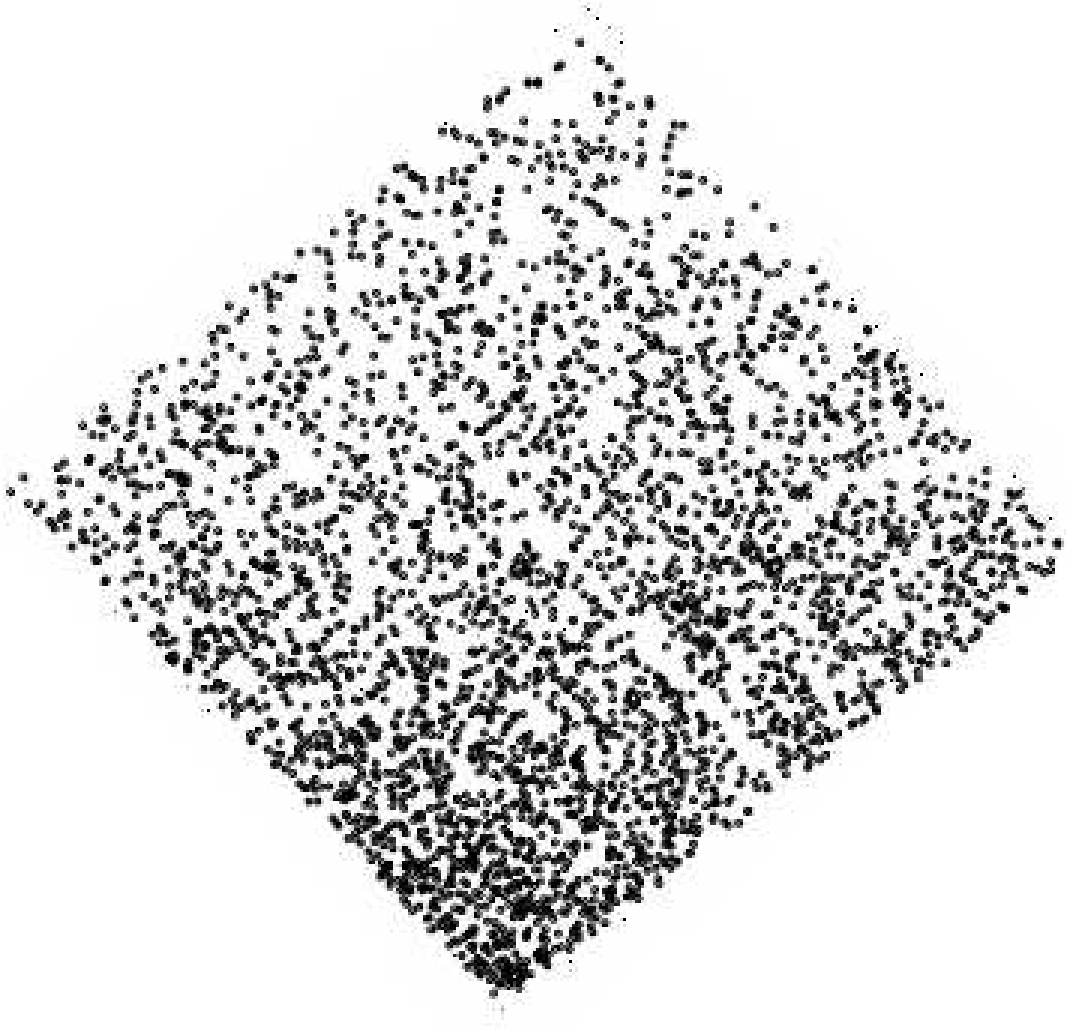}
\caption{The spatial location of the excess field population has a clear
gradient towards the direction of the interacting galaxy pair M60/NGC4647.
This figure plots the positions of the \acsz-based sources with $S/N>5$ and
\acsz$>26.0$ mag, which is the brightness level where the excess starts to
be significant. North is up and East is to left. M60/NGC4647 is to the south
of the field (see also Fig. 1). Note that the gap in the middle is caused
by the gap between the two CCD chips of the ACS/WFC.
}
\end{figure}

\clearpage
\begin{figure}
\plotone{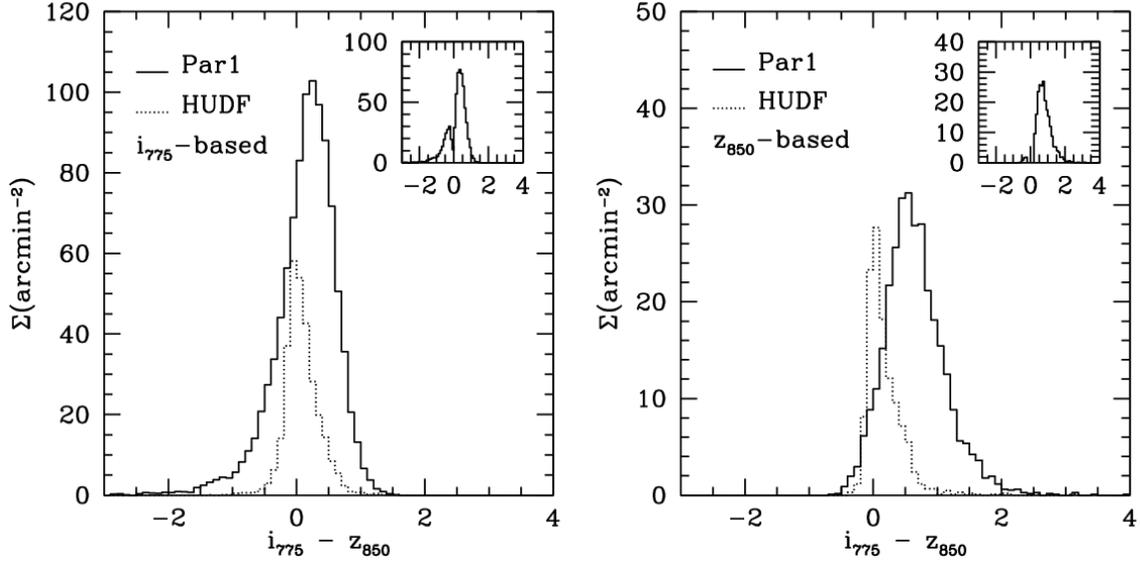}
\caption{The excess population in the parallel fields is red in color. The
solid histograms show the \acsi$-$\acsz\, color distributions of Par1 sources
that are within the magnitude ranges where the excess occurs, while the dotted 
histograms show the distribution of the HUDF sources within the same magnitude
ranges. The insets show the distributions of the net excess using the HUDF as
the reference. The \acsi-based catalogs are used in the left panel, and the
\acsz-based catalogs are used in the right panel. The red wing of the excess
population in \acsz\, continuously extends to the regime where the $i$-dropouts
are selected.
}
\end{figure}

\clearpage
\begin{figure}
\plotone{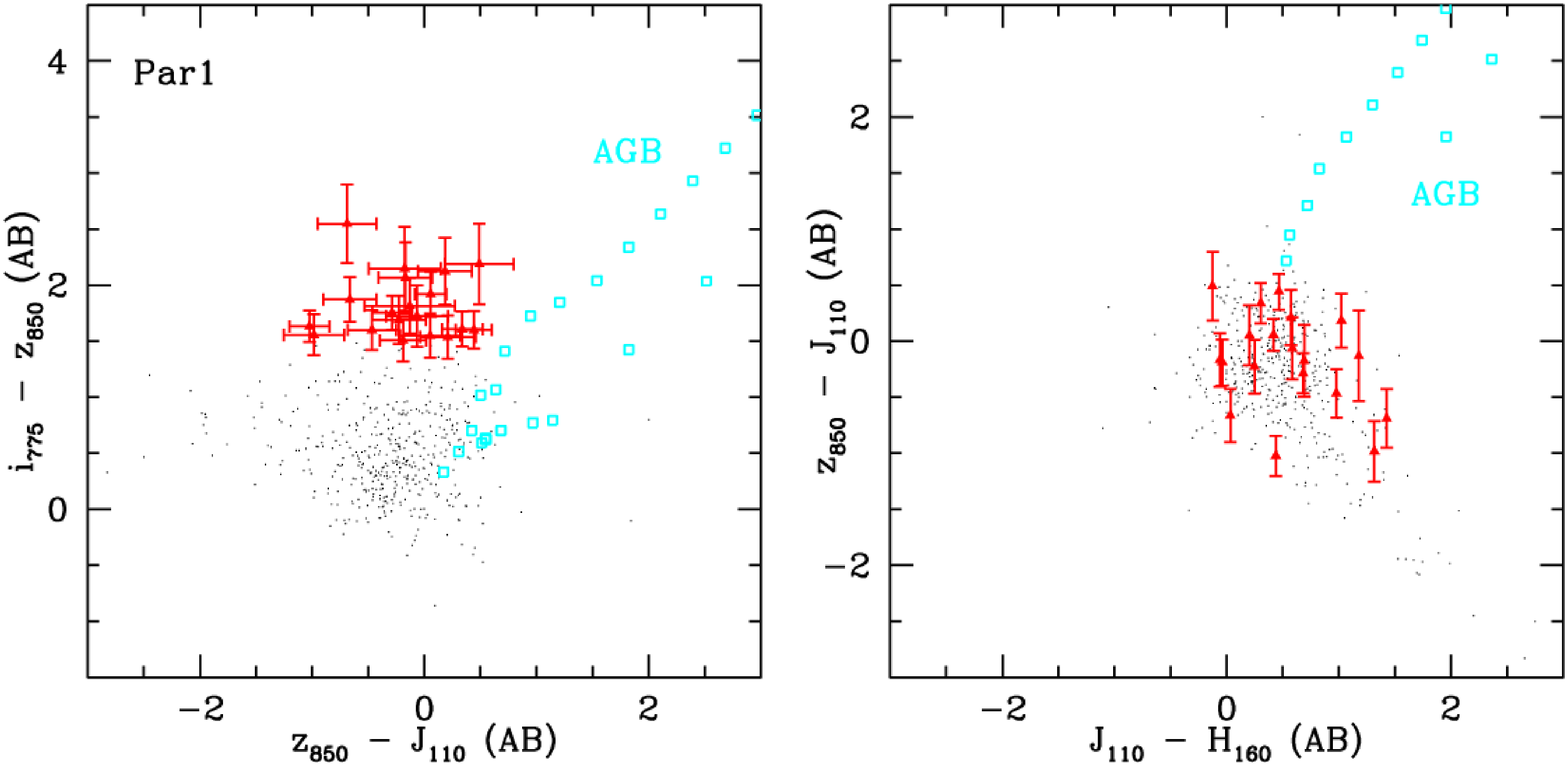}
\caption{While the $i$-dropouts in the two parallel fields might well be
part of the excess stellar population related to M60/NGC4647 instead of
galaxies at $z\approx 6$, we have not yet been able to identify the type of
stars that can create the colors of the $i$-dropouts. The most promising
candidates are AGB stars, which have the necessary luminosities and are red
in colors. However, AGB stars still cannot fully explain the colors of the
$i$-dropouts. To illustrate this, the same color-color diagrams as in the left
panel of Fig. 4 are shown here again, and the colors of AGB stars (cyan
squares) from the models of Lancon \& Mouchine (2002) are superposed. To avoid
confusion, only the models that satisfy \acsi$-$\acsz$>1.3$ mag are superposed
in the right panel.
}
\end{figure}


\begin{thebibliography}{}

\bibitem[]{659} Beckwith, S. V. W., et al. 2006, AJ, 131, 1729
\bibitem[]{660} Bertin, E. \& Arnouts, S. 1996, A\&AS, 117, 393
\bibitem[]{661} Bouwens, R. J., et al. 2003, ApJ, 595, 589
\bibitem[]{662} Bouwens, R. J., et al. 2004, ApJ, 606, L25
\bibitem[]{663} Bruzual, A. G. \& Charlot, S. 1993, ApJ, 405, 538
\bibitem[]{664} Bunker, A. J., Stanway, E. R., Ellis, R. S., \& McMahon, R. G. et al. 2004, MNRAS, 355, 374
\bibitem[]{665} Cohen, S. H., et al. 2003, AJ, 125, 1762
\bibitem[]{666} Dickinson, M., et al. 2004, ApJ, 600, L99
\bibitem[]{667} Durrell, P. R., et al. 2002, ApJ, 570, 119
\bibitem[]{684} Ferguson, A. M. N., et al. 2002, AJ, 124, 1452
\bibitem[]{668} Ferguson, H. C., Tanvir, N. R. \& von Hippel, T. 1998, Nature, 391, 29
\bibitem[]{669} Ford, H., et al. 2003, in ``Future EUV and UV Visible Space Astrophysics Missions and Instrumentation'', eds. J. C. Blades \& O.H. Siegmund, Proc. SPIE, Vol. 4854, 81 
\bibitem[]{670} Fruchter, A. S. \& Hook, R. N. 2002, PASP, 114, 144
\bibitem[]{671} Koekemoer, A. M., Fruchter, A. S., Hook, R. N., \& Hack, W. 2002, in ``The 2002 HST Calibration Workshop : Hubble after the Installation of the ACS and the NICMOS Cooling System'', eds. Santiago Arribas, Anton Koekemoer, and Brad Whitmore, Baltimore, USA, p.337
\bibitem[]{672} Lancon, A. \& Mouhcine, M. 2002, A\&A, 393, 167
\bibitem[]{673} Madau, P. 1995, ApJ, 441, 18
\bibitem[]{674} Pavlovsky, C., et al. 2005, "ACS Data Handbook", Version 4.0, Baltimore: STScI.
\bibitem[]{692} Shang, Z., et al. 1998, ApJ, 504, L23
\bibitem[]{675} Somerville, R. S., et al. 2004, ApJ, 600, L171
\bibitem[]{676} Stanway, E. R., Bunker, A. J., \& McMahon, R. G. 2003, MNRAS 342,439
\bibitem[]{677} Sparks, W. B., et al. 2001, ACS Default (Archival) Pure Parallel Program (ISR01-06) 
\bibitem[]{678} Spergel, D. N., et al. 2003, ApJS, 148, 175
\bibitem[]{679} Spergel, D. N., et al. 2007, ApJS, 170, 377
\bibitem[]{680} Steidel, C. C. \& Hamilton, D. 1992, AJ, 104, 941
\bibitem[]{699} Taylor, V. A., Jansen, R. A., Windhorst, R. A., Odewahn, S. C., \&
Hibbard, J. 2005, ApJ 630, 78
\bibitem[]{681} Williams, B. F., et al. 2007a, ApJ, 654, 835
\bibitem[]{682} Williams, B. F., et al. 2007b, ApJ, 656, 756
\bibitem[]{683} Yan, H., Windhorst, R. \& Cohen, S. 2003, ApJ, 585, L93 (YWC03)
\bibitem[]{684} Yan, H. \& Windhorst, R., 2004, ApJ, 600, L1 (YW04)
\bibitem[]{705} Zackrisson, E., et al. 2006, ApJ, 650, 812
\bibitem[]{706} Zheng, Z., et al. 1999, AJ, 117
\bibitem[]{707} Zibetti, S. \& Ferguson, A. M. N., 2004, MNRAS, 352, L6
\bibitem[]{708} Zibetti, S., White, S. D. M \& Brinkmann, J. 2004, MNRAS, 347, 556

\end{thebibliography}
\end{document}